\documentclass[12pt]{iopart}
\usepackage{iopams}
\usepackage{graphicx}
\begin{document}

\title[Particle confinement and quantum computing]{Many-particle confinement
by constructed disorder and quantum computing}

\author{M I Dykman\dag, L F Santos\dag\ddag\ and M Shapiro\S  }

\address{\dag\ Department of Physics and Astronomy
and Institute for Quantum Sciences,
Michigan State University, East Lansing, MI 48824 USA}

\address{\ddag\ Department of Physics and Astronomy,
Dartmouth College,
Hanover, NH 03755 USA }

\address{\S\ Department of Mathematics
and Institute for Quantum Sciences,
Michigan State University, East Lansing, MI 48824 USA}

\ead{dykman@pa.msu.edu, Lea.F.Dos.Santos@Dartmouth.edu, mshapiro@math.msu.edu}

\begin{abstract}
Many-particle confinement (localization) is studied for a 1D system of
spinless fermions with nearest-neighbor hopping and interaction, or
equivalently, for an anisotropic Heisenberg spin-1/2 chain. This
system is frequently used to model quantum computers with perpetually
coupled qubits. We construct a bounded sequence of site energies that
leads to strong single-particle confinement of all states on
individual sites. We show that this sequence also leads to a
confinement of {\it all} many-particle states in an {\it infinite}
system for a time that scales as a high power of the reciprocal
hopping integral. The confinement is achieved for strong interaction
between the particles while keeping the overall bandwidth of site
energies comparatively small. The results show viability of quantum
computing with time-independent qubit coupling.

\end{abstract}




\maketitle

\section{Introduction}


Spatial localization of excitations is central to quantum control
and measurement. It is often required both for addressing
excitations individually and for determining their states.
Localization is particularly important for quantum computing. A
quantum computer (QC) is usually thought of as a system of physical
objects that can be modeled by two-level systems, qubits. The
objects can be individually accessed to perform quantum operations.
They also provide a natural physical basis for measurement.  In many
proposed implementations of QCs the qubit-qubit interaction is not
turned off \cite{liquid_NMR}--
\cite{Devoret_review04}. Generally, the interaction leads to hopping
of excitations between the qubits. The hopping is not related to
relaxation, but they both may be devastating for QC
operation. Therefore eliminating excitation hopping, or at least
significantly slowing down the hopping rate, is a prerequisite for
quantum computation.

The problem of localization has a long history in physics, most
notably in condensed matter physics, where localization is due to
broken translational symmetry. Here it is usually discussed for
particles. The relation to the case of qubit excitations is
particularly simple for one-dimensional systems. The presence of
a particle on site $n$ corresponds to the $n$th qubit being excited,
and the particle energy is the excitation energy. Much work has been
done on single-particle localization. One of the best known examples
is Anderson localization due to random disorder \cite{Anderson}, which
has been studied in depth and by now is well understood.  It has been
demonstrated also, both theoretically and experimentally
\cite{Thouless83}--
\cite{Merlin85}, that
single-particle localization occurs also in a different type of
systems, where the symmetry breaking is due to an incommensurate
periodic potential.

The work on localization in systems with many excitations is more
recent \cite{many}. It has demonstrated that the particle-particle
interaction can significantly affect localization \cite{Basko05}. We note
that numerical studies of this problem are not very helpful,
because the Hilbert space of a many-particle system is exponentially
large, with the exponent proportional to the number of particles, and
therefore only systems with a small number of particles can be
simulated with classical computers.

In the present paper we are interested in a specific formulation of
the problem of localization. We study a many-particle (or
equivalently, many-excitation) system and consider localization of
{\em all} states in such system. More than that, bearing in mind
applications to quantum control and quantum computing, we study strong
on-site localization, where the wave function of each particle is
mostly confined to one site. This is a stronger requirement than
exponential decay of the wave functions at large distances. Formally
it corresponds to the localization length being less than the
intersite distance, for all states. We call such localization {\it
on-site confinement} or simply confinement, for brevity.

Many-particle confinement does not arise in a disordered system with
bounded random site energies \cite{Shepelyansky}. Indeed, consider a
chain where particles occupy $N$ sites. For short-range hopping an
$N$-particle state is directly coupled to $\sim N$ other $N$-particle
states. With probability $\propto N$ one of them will be in resonance
with the initial state, provided the site energies are uniformly
distributed over a finite-width band. This means that the energy
difference between the two states will be less or of the order of the
intersite hopping integral $J$ (we set $\hbar = 1$).  Resonating
states are hybridized over time $\sim J^{-1}$.

In contrast to random disorder, in a QC the disorder can often be
constructed at will, because site excitation energies can be
individually controlled. This makes it possible to look at
many-particle localization from a new perspective. Not only is
``localization by construction'' necessary for quantum computing, but
in itself it is an exciting area of applications of QCs, which is of
special interest to condensed-matter physics. This is because the very
possibility of localization of all stationary many-particle states is
an open question, as the distance between the energy levels is
exponentially small for a large number of particles.

In what follows we extend the previous work \cite{DISS04} to show
that, for an {\em infinite} 1D system, many-particle confinement can
be obtained by constructing a sequence of site energies. The proposed
sequence has a comparatively narrow bandwidth. This is physically
reasonable and is also important for applications in quantum
computing, because the qubit tuning range is limited. Additionally, in
a QC a smaller bandwidth is desirable for a higher speed of quantum
gate operations.

Rather than studying the stationary states, we put emphasis on the
{\it localization lifetime}, i.e., the time during which all
many-particle states remain confined to the initially occupied
sites. We show that this lifetime can largely exceed $J^{-1}$. We also
find that, for moderately long sections of the constructed energy
sequence, all stationary many-particle states can be strongly
localized.

The idea of our approach is to map the problem of the lifetime of
localized states on the problem of many-particle scattering. We first
construct site energies that lead to strong confinement of all
single-particle states. Then delocalization of many-particle states
corresponds to resonant scattering in which, as a result of the
interaction, particles move away from the initially occupied
sites. For real scattering to happen, the final configuration must be
close in energy to the initial configuration. The lifetime of a state
is proportional to the inverse of the matrix element of hopping onto a
resonating state.

A large localization lifetime is obtained if resonant scattering
between many-particle configurations requires several virtual
single-particle transitions via nonresonant states. The amplitudes of
such transitions are small.  The lifetime is determined by their
minimal number, which we call the ``order'' of the transition. When it
is limited to 4, it is sufficient to consider transitions that involve
no more than 4 particles. The approach developed in this paper allows
us to obtain long lifetime not for comparatively weak \cite{DISS04}
but rather for strong interparticle coupling. As we show, this can be
done without a significant increase of the bandwidth of site energies
compared to the weak-coupling case. The strong-coupling confinement is
of interest, in particular, for some models of QC's, like a QC based
on electrons on helium, where the dimensionless coupling parameter
$\Delta$ may be as large as $\sim 10$ \cite{Parmoon}.

The paper is organized as follows. In Section~II a qualitative picture
of localization is provided. Section~III contains a description of the
effective localizing sequence \cite{DISS04} and a discussion of
single-particle localization. In Sec.~IV the localization lifetime is
introduced and many-particle scattering is described. In Section~V the
bands of combination energies for many-particle transitions are
analyzed and the conditions for eliminating resonances for transitions
up to 4th order are found. Section~VI contains a brief discussion of
the results.

\section{Localization by constructed disorder: preliminaries}

Finding the ways of localizing excitations has been appreciated as an
important problem of quantum computing. A natural formulation here is
to map a QC onto a system of interacting spins, with $S=1/2$. The
spins are in a magnetic field, and the qubit excitation energy is the
spin Zeeman energy. Several approaches to solving the localization
problem have been put forward. An interesting idea substantiated by
extensive numerical calculations is to arrange the Zeeman (site)
energies into a ladder, so that none of them is in resonance with each
other \cite{Berman-01}. The bandwidth of site energies in this case
linearly increases with the size of the system, and localization was
studied for the Ising interaction between the spins.  Another
interesting idea is to create interaction-free subspaces
\cite{Zhou02}. The qubits that form these subspaces are separated from
each other by the isolators, which are qubits prepared in a special
state. The bandwidth of site energies remains finite in this case even
for an infinite system, and the spin-spin interaction is not limited
to the Ising one. However, the preparation requires that initially the
interaction between the qubits is turned off.

We propose a different approach, largely inspired by the results on
Anderson localization in condensed-matter physics. We show that, by
constructing site disorder, confinement can be achieved for a finite
energy bandwidth and for the nearest-neighbor interaction that has a
general form and is never turned off.

\begin{figure}[h]
\begin{center}
\includegraphics[width=2.5in]{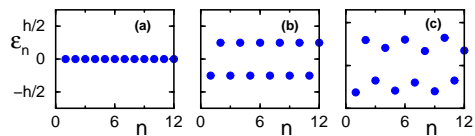}
\end{center}
\caption{The successive detuning of site energies $\varepsilon_n$ in
order to achieve single-particle localization in a chain. (a) All
energies are the same, $\varepsilon_n=0$. (b) Nearest neighbors are
detuned from each other by an energy $h$. (c) Second, fourth, and
higher-order neighbors are detuned from each other, with the detuning
that becomes smaller and smaller with the increasing inter-site
distance. The data in (b) and (c) correspond to the energy sequence
(\ref{sequence}) with $\alpha =0$ and $0.1$, respectively.}
\label{fig:detuning}
\end{figure}

Constructing a sequence of site energies $\varepsilon_n$ that
leads to single-particle (single-excitation) confinement is quite
straightforward. If all $\varepsilon_n$ are the same, as in
Fig.~\ref{fig:detuning}(a), intersite hopping leads to formation of a
band of fully delocalized states, with energy bandwidth proportional
to the hopping integral $J$.

The first step towards localization is to detune neighboring sites
from each other by an energy $h\gg J$, see
Fig.~\ref{fig:detuning}(b). In this case nearest neighbor hopping
becomes nonresonant. However, this is insufficient for localization,
even though in quantum computing literature there are claims to the
opposite. Indeed, a particle can make a virtual transition to a
neighboring site and from there to the resonating next nearest
site. The corresponding transition matrix element is
$J^2/h$. Therefore the states will be again band-like, with two bands
of width $\propto J^2/h$, which are centered at $\pm h/2$.

A natural next step is to detune the energies of 2nd neighbors. The
detuning in this case should again exceed the matrix element of
effective hopping. However, since the latter is $\sim J^2/h\ll J$, the
detuning can be smaller than $h$ by a factor $\sim J/h$, see
Fig.~\ref{fig:detuning}(c). At the next step it is necessary to detune
4th neighbors, but here the detuning can be smaller by an extra factor
$J/h$. The process should be continued. It leads to a geometric-series
type level detuning, with a finite overall bandwidth of the energy
spectrum. A specific example is discussed below.

A significantly more complicated situation arises for a many-particle
system. Here, one should tune away from each other not only site
energies, but also their combinations. This is necessary in order to
eliminate resonant multi-particle transitions. The total number of
states for $N$ particles on an $L$-site long chain is $L!/N!(L-N)!
\propto 4^N$ for $N=L/2$, and therefore for large $N$ the distance
between the energy levels is exponentially small. This makes the
problem of eliminating resonances hard. Nevertheless, as mentioned
above, one can eliminate resonances that require at least a certain
minimal number of virtual transitions to nonresonant states, leading
to a long time over which all many-particle configurations remain
confined to their sites.

\section{Localization of stationary states}

\subsection{One-parameter sequence of site energies}

A 1D spin system or a qubit chain can be mapped onto a system of
spinless fermions via the Jordan-Wigner transformation
\cite{Jordan_Wigner}. The Hamiltonian of the fermions is
\begin{eqnarray}
\label{hamiltonian_fermions}
H=H_0+ H_i,\\
H_0=\sum\nolimits_n\varepsilon_na_n^{\dagger}a_n+ {1\over
2}J\sum\nolimits_n\bigl( a_n^{\dagger}a_{n+1}+a_{n+1}^{\dagger}a_n\bigr),
\nonumber\\
H_i= J\Delta\sum\nolimits_n
a_n^{\dagger}a_{n+1}^{\dagger}a_{n+1}a_n.\nonumber
\end{eqnarray}
Here, $a^{\dagger}_n, a_n$ are the fermion creation and annihilation
operators.  The site fermion energies $\varepsilon_n$ are the Zeeman
energies of the spins or the excitation energies of the qubits.

The spin-spin interaction leads to two effects, in terms of
fermions. One is the intersite fermion hopping, with the hopping
integral $J$. It is determined by the $XY$ spin-spin interaction
$S_n^xS_{n+1}^x+ S_n^yS_{n+ 1}^y$ or, in the case of qubits, by the
interaction between the qubit transition dipoles. Fermion hopping is a
single-particle effect: the $XY$ spin-spin interaction does not lead
to interaction between the fermions. A different role is played by the
Ising spin-spin interaction $S_n^zS_{n\pm 1}^z$, or, in the case of
qubits, by the change of the transition energy of a qubit depending on
whether or not the neighboring qubit is excited. It leads to the
interaction of fermions, which is described by the Hamiltonian
$H_i$. The parameter $\Delta$ characterizes the anisotropy of the
spin-spin interaction: for the isotropic Heisenberg model $\Delta
=1$. For concreteness we set $J,\Delta > 0$.

A simple bounded sequence of site energies that implements the
construction described in Sec.~2 for a semi-infinite chain with $n\geq
1$ has the form \cite{DISS04}
\begin{equation}
\label{sequence}
\varepsilon_n={1\over 2}h\left[(-1)^n -
\sum\nolimits_{k=2}^{n+1}(-1)^{\lfloor
n/k\rfloor}\alpha^{k-1}\right],
\end{equation}
where $\lfloor \cdot\rfloor$ is the integer part.

Besides the energy scale $h$, sequence (\ref{sequence}) is
characterized by one dimensionless parameter $\alpha < 1$. The overall
energy bandwidth is $W\lesssim h/(1-\alpha)$. For $\alpha\ll 1$ the
energies {\ref{sequence}) form two subbands centered at $\pm h/2$.
Numerically, the subbands remain well separated for $\alpha < 0.4$.

\subsection{Single-particle localization}

The expression (\ref{sequence}) has a special symmetry: as $n$
increases, the coefficients at any given power $\alpha^q$ are repeated
with period $2(q+1)$.  This symmetry was used to obtain an analytical
expression for the amplitude $K_n(m)$ of a single-particle transition
from site $n$ to site $n+m$ \cite{DISS04}. The amplitude $K_n(m)$
quasi-exponentially decays for all states in a semi-infinite
chain. For small $\alpha$
\begin{equation}
\label{result}
K_n(m)= \alpha^{-\nu m}\,(J/2h)^{m}.
\end{equation}
The exponent $\nu$ is determined by $\lim_{m\to\infty}m^{-1}\, \log
[K_n(m)]$. It is not the reciprocal average decay length: it varies
from site to site, $\nu\equiv \nu(n)$, as shown in
Fig.~\ref{fig:exponents}.
\begin{figure}[h]
\begin{center}
\includegraphics[width=2.5in]{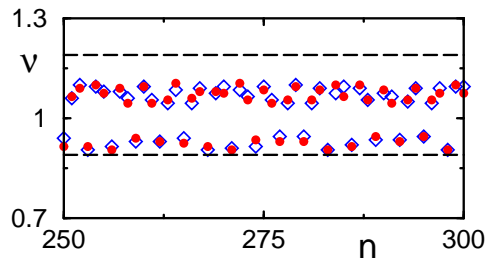}
\end{center}
\caption{The exponent $\nu=\nu(n)$ of the $\alpha$ dependence of the
transition amplitude $K_n(m)$ \protect(\ref{result}) as a function of
site $n$. The data refer to the distances $m=200$ (diamonds) and
$m=-200$ (circles). The dashed lines show the analytical limits on
$\nu$ for $m\to\infty$. The values of $\nu$ for the same site but for
positive and negative $m$ are different but lie within the same
bounds.}
\label{fig:exponents}
\end{figure}

The values of $\nu$ are bounded to a narrow region centered at $\nu=
1$, with $0.89 < \nu < 1.19$. For estimates one can use $\nu=1$, i.e.,
\begin{equation}
\label{exp_decay}
K_n(m)\approx K^{m},\qquad K= J/2\alpha h.
\end{equation}

Along with decay ``forward'', i.e., for $m>0$ \cite{DISS04}, we have
studied decay ``backward'', i.e., for $m<0$. Because of the
symmetry of the sequence we expect that the decay backward should be
also quasi-exponential in $m$, and the bounds on $\nu$ should be the same as
for positive $m$. The numerical data in Fig.~\ref{fig:exponents}
substantiate this conjecture. Therefore in the estimates given below
we use Eq.~(\ref{exp_decay}) in the form $K_n(m)\approx K^{|m|}$,
which applies to both positive and negative $m$.

The characteristic decay length of the transition amplitude is $1/|\ln
K|$. For $K\ll 1$, i.e. $J\ll 2\alpha h$, all single-particle states
are strongly localized and essentially confined to single sites, with
small tails on neighboring sites.

\subsection{Inverse participation ratio}

The extent to which stationary states in a system are confined can be
conveniently characterized by the inverse participation ratio
(IPR). For an $N$-particle eigenstate $|\psi_{N\lambda}\rangle$
($\lambda$ enumerates the eigenstates) it is given by
\begin{equation}
\label{IPR}
I_{N\lambda}=\left(\sum\nolimits_{n_1<\ldots<n_N}\bigl\vert\langle
\Phi_{n_1\ldots n_N}|\psi_{N\lambda}\rangle\bigr\vert^4\right)^{-1},
\end{equation}
where $|\Phi_{n_1\ldots n_N}\rangle$ is an on-site $N$-particle wave
function (quantum register) with particles on sites $n_1,\ldots,n_N$.

If the stationary state $|\psi_{N\lambda}\rangle$ is fully localized,
all terms in the sum (\ref{IPR}) are equal to $0$ except one, which is
equal to $1$, and $I_{N\lambda}=1$. In the opposite case of
delocalized eigenstates all matrix elements in the sum (\ref{IPR}) are
small, and $I_{N\lambda}\gg 1$ for a large system. The on-site
confinement that we are interested in corresponds to $I_{N\lambda}$
being close to $1$ for {\it all} states $\lambda$.

\begin{figure}[h]
\begin{center}
\includegraphics[width=3.0in]{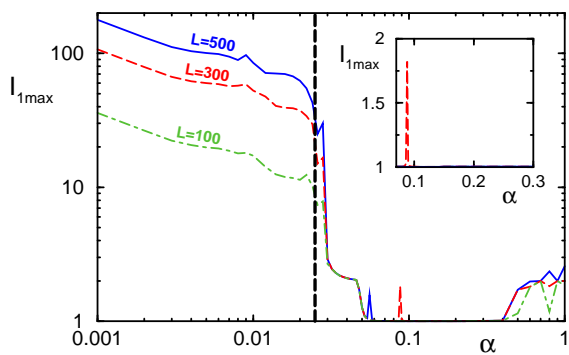}
\end{center}
\caption{The maximum inverse participation ratio $I_{N\,\max}$ for one
particle ($N=1$) in an $L$-site chain.  The energies $\varepsilon_n$
are given by Eq.~(\ref{sequence}), and $h/J=20$. The data refer to the
segments of the chain with $1\leq n\leq L$. The plateau and sharp
isolated peaks in the region $0.03\lesssim \alpha < 0.1$ result
primarily from pairwise hybridization of states. An example of the
peak for $L=300$ is shown in the inset on the linear in $\alpha$
scale. In the whole region $0.1 < \alpha \lesssim 0.4$ the IPR
$I_{1\,\max} \approx 1.003$, which indicates strong confinement of all
stationary states. The asymptotic localization threshold $\alpha
=J/2h$ is shown by the dashed line.}
\label{fig:IPR_weak}
\end{figure}

In Fig.~\ref{fig:IPR_weak} we show the results of the numerical
calculation of the maximal IPR
$I_{N\,\max}=\max_{\lambda}I_{N\lambda}$ for energy sequence
(\ref{sequence}) for one particle, $N=1$. In agreement with
Eqs.~(\ref{result}), (\ref{exp_decay}), all single-particle states are
strongly localized for $\alpha\geq 0.1$, where $K \leq 0.25$. It is
seen from Fig.~\ref{fig:IPR_weak} that $I_{N\,\max}$ sharply drops
down for $\alpha\approx 0.03$, which is close to the asymptotic value
$J/2h=0.025$ given by the condition $K=1$. For $0.03\lesssim
\alpha\lesssim 0.05$ the IPR is close to 2 for all $L$, which
indicates that only a few on-site states are appreciably hybridized;
for example, $I_{1\lambda}=2$ if two on-site states are hybridized
with equal weight in the wave function $|\psi_{1\lambda}\rangle$.

Depending on the chain length, pairwise hybridization also
occasionally occurs for particular values of $\alpha$ in the range
$0.05 \lesssim\alpha < 0.1$. An example is the peak for $L=300$ shown
in the inset of Fig.~\ref{fig:IPR_weak}. It is due to hybridization of
the states on sites 296 and 300 \cite{DISS04}. The hybridization
happens because the interaction $Ja^{\dagger}_na_{n+1}$ shifts the
site energies by $\sim J^2/h$. The shift is different for the state on
the boundary, since it has less neighbors. In the particular case of
the chain with $L=300$ a simple calculation shows that this difference
leads to resonance between the aforementioned on-site states. For the
chosen $J/2h$ in the region $\alpha \lesssim 0.06$ the wave functions
have longer tails than what follows from the asymptotic expression
(\ref{result}), and therefore the states away from the boundary are
still affected by it. This leads to the peak of $I_{N\,\max}$ for
$L=500, \alpha=0.056$ seen in Fig.~\ref{fig:IPR_weak}. Changing $L$ in
the region $\alpha \lesssim 0.06$ leads to disappearance of some peaks
and onset of other peaks, with hybridization of different states. It
is important, however, that for $0.4\gtrsim \alpha \gtrsim 0.1$ these peaks
disappear and all states are strongly localized.

We have also studied \cite{DISS04} a chain of length $L=12$ with
$N=L/2$ excitations (where the number of states is maximal, for a
given $L$). For the section of the chain $1\leq n \leq L$, the maximal
IPR $I_{6\,\max}\leq 1.02$ for $h/J=20$, in a broad range $0.2\lesssim
\alpha\lesssim 0.4$, except for sharp isolated peaks for certain
$\alpha$. This indicates strong confinement of all many-particle
states. However, the IPR for large $L$ and $N$ could not be studied
because of the large number of states.


\section{The localization lifetime}

Confining a many-particle system is much more complicated than a
single-particle one. It requires eliminating all possible
many-particle combinational resonances $\varepsilon_{n_1}+\ldots
+\varepsilon_{n_k}\approx \varepsilon_{m_1}+\ldots
+\varepsilon_{m_k}$. Therefore we approach the problem of strong
many-particle confinement from a different perspective. Rather than
studying stationary states, we will consider the time over which
interacting particles leave the sites where they were placed
initially. We call this time the localization lifetime, $t_{\rm loc}$.

The decay of on-site states is not exponential, at least for not too
long times, and the time $t_{\rm loc}$ is not the decay time.  On-site
states evolve primarily due to resonant transitions to other on-site
states with the same energy. The transition rates between different
resonating many-particle states vary strongly provided the
single-particle states are all strongly localized. First there occurs
hybridization of a given on-site many-particle state with the
resonating state that is most strongly coupled to it. Then these
states may further hybridize with other resonating states, but this
may take much longer. For our purpose, the localization lifetime is
the time of the first hybridization. In other words, $t_{\rm loc}$ is
the reciprocal maximal rate of a transition to a resonating on-site
state.

An important characteristic of the dynamics of quantum systems is the
coherence time $\sim t_{\rm coh}$. In any real system $t_{\rm coh}$ is
finite due to thermal fluctuations, decay into excitations of the
medium, and fluctuations induced by external noise. We are interested
in strong coherent confinement, in which case $t_{\rm loc}$ calculated
in the neglect of decoherence is $\gtrsim t_{\rm coh}$. We note that,
in fact, decoherence can increase the time particles spend on their
sites by suppressing coherent quantum transitions; however, this
suppression is outside the scope of the present paper.

In systems proposed for quantum computing $J$ and $J\Delta$
characterize the rate of two-qubit operations, and typically the
number of operations that can be performed over the coherence time is
$\lesssim 10^5$. Therefore, for practical purposes, in a quantum
computer excitations will be localized provided
%
$t_{\rm loc}$ exceeds $J^{-1}$ or $(J\Delta)^{-1}$ by a factor $10^5$.
%
This condition must be satisfied for {\it all} many-excitation states
and should apply to an {\it infinite} system for the quantum computer
to be scalable.

\subsection{Many-particle scattering}

Even though the intersite transitions themselves are single-particle,
resonant many-particle transitions may happen even where single-particle
resonances are eliminated. This is a consequence of the energy
exchange between interacting particles and also the energy change due
to the interaction.  The interaction (\ref{hamiltonian_fermions}) is
nearest-neighbor two-particle. Therefore a many-particle transition
results from a sequence of single-particle intersite transitions and
scattering of particles on neighboring sites. In what follows we
assume that, even though the particle-particle coupling may be strong,
$\Delta \gg 1$, still $\Delta \ll h/J$. Then nonresonant many-particle
effects can be considered by perturbation theory.

A simple example of a Feynman diagram that describes a two-particle
transition is shown in Fig.~\ref{fig:diagram}. In this example
particles located initially on sites $(n,n+3)$ first make two
single-particle transitions $(n,n+3)\to (n,n+2) \to (n,n+1)$, then
experience scattering by each other and exchange their energies, and
then make two more single-particle transitions $(n,n+1)\to (n,n+2) \to
(n-1,n+2)$. The overall sequence of transitions will be resonant if
$\varepsilon_{n}+ \varepsilon_{n+3}\approx
\varepsilon_{n-1}+\varepsilon_{n+2}$.

An alternative approach to the analysis of many-particle scattering is
based on diagonalizing the single-particle part of the Hamiltonian
(\ref{hamiltonian_fermions}) \cite{DISS04}. Since all single-particle
states are strongly confined, the resulting states are very close to
on-site states. However, the particle-particle interaction acquires a
small off-diagonal long-range part. It is clear in this approach that
it is the interaction that leads to resonant many-particle scattering.

\begin{figure}[h]
\begin{center}
\includegraphics[width=3in]{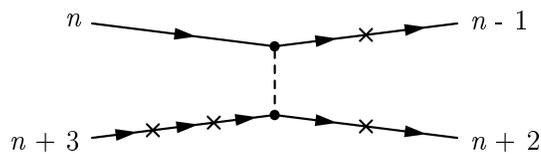}
\end{center}
\caption{A diagram for a two-particle transition from sites $(n,n+3)$
to $(n-1,n+2)$. The crosses indicate single-particle transitions to
neighboring sites, and the dashed line indicates scattering of two
particles located on neighboring sites.}
\label{fig:diagram}
\end{figure}

>From inspection of the diagram in
Fig.~\ref{fig:diagram} it is obvious that, for a many-particle transition to
occur, the particles have to be brought to neighboring sites. This
requires single-particle transitions via nonresonant sites. For a
two-particle resonant transition $(k_4,k_3)\leftrightarrow (k_1,k_2)$
(we choose for concreteness $k_3>k_4$ and $k_2>k_1$) the minimal
number of the needed single-particle transitions is given by the
parameter
\begin{equation}
\label{varkappa}
\varkappa=\min_p(|k_1-p|+|k_2-p-1|+|k_3-p-1|+|k_4-p|).
\end{equation}
This is easy to see: the particles move from the initially occupied
sites to neighboring sites $(p,p+1)$ and from there to the final
sites; the minimum over $p$ in Eq.~(\ref{varkappa}) is taken so as to
minimize the total distance involved in the transition.

For large $\varkappa$, the overall amplitude of the involved
single-particle transitions is $K^{\varkappa}$, according to
Eq.~(\ref{exp_decay}). Since the two-particle interaction energy is
$J\Delta$, the localization lifetime of the state $(k_4,k_3)$ with
respect to a transition to the state $(k_1,k_2)$ as given by the
reciprocal transition matrix element is
\begin{equation}
\label{t_loc}
t_{\rm loc}\approx \left(J\Delta K^{\varkappa}\right)^{-1}, \qquad
\varkappa \gg 1.
\end{equation}
This estimate is made assuming that the transition involves only one
particle-particle collision. The role of multiple collisions will be
studied in a separate paper \cite{DS_to_be}.

We are interested in the localization lifetime with respect to all
resonant transitions and for all states. Then, from Eq.~(\ref{t_loc}),
$\varkappa$ should exceed a certain minimal value $\varkappa_{\min}$
for all resonant transitions. This means that all resonances separated
by less than $\varkappa_{\min}$ steps must be eliminated.

In what follows we will show explicitly how, in an infinite system,
all resonances with $\varkappa \leq 4$ can be eliminated for
moderately large $h/J\ll 10^2$ even for $\Delta$ as large as $\Delta
\sim 10$. Then, with $K\sim 0.1$, the single-collision estimate
(\ref{t_loc}) leads to $t_{\rm loc} \gtrsim 10^5(J\Delta)^{-1}$.

The important simplification from considering resonances with
$\varkappa \leq 4$ is that the analysis can be limited to transitions
in which only two, three, and four particles change their sites. We
leave the analysis of three- and four-particle transitions for a
separate paper \cite{DS_to_be}; in the present paper we will only give
its result and concentrate on eliminating resonances related to
two-particle transitions.

The change of the on-site energy in a two-particle transition
$(k_4,k_3)\leftrightarrow (k_1,k_2)$ is
\begin{equation}
\label{pair_diff}
\delta\varepsilon
=|\varepsilon_{k_1}+\varepsilon_{k_2}-
\varepsilon_{k_3}-\varepsilon_{k_4}|.
\end{equation}
To avoid resonances, this energy difference should largely exceed the
transition matrix element. However, this is not enough, because of the
change of the interaction energy that may accompany the transition. To
the lowest order in $J/h$ it is determined by the change of the number
of nearest neighbors of the hopping particles. The analysis of
diagrams of the type shown in Fig.~\ref{fig:diagram} shows that, for
$\varkappa\leq 4$, this number may remain unchanged or change by one
or two, in which case the interaction energy will change by $J\Delta$
or $2J\Delta$.

In the previous work \cite{DISS04} we studied the case $\Delta
\lesssim 1$. In this case it was sufficient to open a gap at $\delta
\varepsilon =0$ with width $>2J\Delta$. However, numerically the gap
was relatively narrow, it was limited to $\delta\varepsilon/h \sim
0.01$. Therefore a direct application of the results to the case of
large $\Delta$ would require using very large $h/J$, which is
undesirable. In the next Section we develop an alternative approach
that applies in the strong-interaction case.

\section{The spectrum of two-particle transition energies}

The overall structure of the differences of site energies
for two-particle transitions $\delta\varepsilon$ is fairly simple. If
$\alpha=0$ in Eq.~(\ref{sequence}), we have $\varepsilon_n=\pm
h/2$. Then from Eq.~(\ref{pair_diff}) $\delta\varepsilon$ takes on
three values, $\delta\varepsilon=0,h,2h$. With increasing $\alpha$
each of these values splits into a band. For small $\alpha$ the
bandwidth is $\propto\alpha h$ and the bands are well separated.

On the other hand, the change of the interaction energy in a
two-particle transition is close either to zero or $J\Delta$ or
$2J\Delta$, as explained above. To avoid resonances related to this
energy change, it should lie within the gaps of $\delta \varepsilon$.

A natural way to obtain strong confinement while keeping the bandwidth
$h$ of site energies moderately large is the following. All
transitions where the interaction energy change is close to zero
should have $\delta\varepsilon$ that is larger than this change but
may be still much smaller than $h$, see below. On the other hand, all
transitions where the interaction energy change is $J\Delta$ or
$2J\Delta$ should have $\delta\varepsilon$ in the gap centered at
$h/2$. Then the ratio $J\Delta/h$ does not have to be extremely small.
Even though this is not the only option, we found that it allows using
not too small values of $\alpha$, which is advantageous in terms of
robustness with respect to errors in the site energies. The inequality
$J\Delta/h < 1/2$ imposes a much less restrictive condition on the
bandwidth $h$ than the requirement that $2J\Delta$ is smaller than the
gap of $\delta\varepsilon$ at zero energy \cite{DISS04}.

We denote the widths of the gaps of $\delta \varepsilon$ at energies
$0, J\Delta$, and $2J\Delta$ by $\delta_{0}$, $\delta_1$, and
$\delta_2$, respectively. These widths should exceed the matrix
elements of the appropriate transitions. They should also be larger
than the energy renormalization due to (possible) occupation of the
second neighbors of the sites involved in a transition.  The
renormalization can be obtained by noticing that the energy change is
due to making a virtual transition on a neighboring site, interacting
with the particle on the next neighboring site, and making a
transition back. This gives the energy shift $\sim J\Delta
(J/2h)^2$. Therefore we need
\begin{equation}
\label{gap_condition}
\delta_{0,1,2}\gg J^3\Delta/(2h)^2.
\end{equation}

\subsection{Zero-energy gap}

Meeting the condition (\ref{gap_condition}) for the zero-energy gap
$\delta_{0}$ requires a modification of the site energy sequence
(\ref{sequence}).  The most ``dangerous'', for $\varkappa \leq 4$, is
the resonance in $\delta\varepsilon$ associated with a transition
$(n,n+1) \leftrightarrow (n-1,n+2)$ \cite{DISS04}. For this transition
$\varkappa = 2$, whereas the energy difference is $\delta\varepsilon
\sim \alpha^{\xi} h$ with $\xi \geq 4$ for all $n=6k-1$. This is a
very small $\delta\varepsilon$ for $\varkappa=2$, because the
transition matrix element is comparatively large in this case. In
addition we have $\delta\varepsilon \sim
\alpha^{n-1}h$ when $n$ and $n+2$ are prime numbers, which leads to
$\delta_0\to 0$ in an infinite chain.

A simple way to eliminate this resonance and to open a gap $\delta_0$
is to modify energy sequence (\ref{sequence}) by shifting the energy on
each 6th site,
\begin{equation}
\label{correction_6}
\varepsilon_n^{\rm md} = \varepsilon_n +(h/2)\alpha^{\prime} \qquad
{\rm for}\qquad n=6k,
\end{equation}
while keeping it unchanged on other sites, $\varepsilon_n^{\rm md} =
\varepsilon_n$ for $n\neq 6k$ \cite{DISS04}. The parameter
$\alpha^{\prime}$ should be chosen in such a way as to avoid creating
new resonances.

All $\varkappa \leq 4$ transitions with small
$\delta\varepsilon$ can be found by modifying the diagram in
Fig.~\ref{fig:diagram}. The rule is simply that the numbers of even
and odd sites should be the same at the input and output
\cite{DS_to_be}. The result coincides with what was obtained earlier
\cite{DISS04} using a different technique. The zero-energy gap for the
sequence (\ref{correction_6}) is $\delta_0 \gtrsim \alpha^3h$ for
$\alpha \gg \alpha^{\prime}\gg \alpha^3$. This gap is illustrated in
Fig.~\ref{fig:zero_energy} where we plot the change in modified site
energies
\begin{equation}
\label{band_modified}
\delta\varepsilon^{\rm md}_n=|\varepsilon_{k_1}^{\rm md}+
\varepsilon_{k_2}^{\rm md}-
\varepsilon_{k_3}^{\rm md}-\varepsilon_{k_4}^{\rm md}|
\end{equation}
for all $\varkappa\leq 4$ transitions in which one of the
sites $k_{1,2,3,4}$ is $n$.

\begin{figure}[h]
\begin{center}
\includegraphics[width=3.0in]{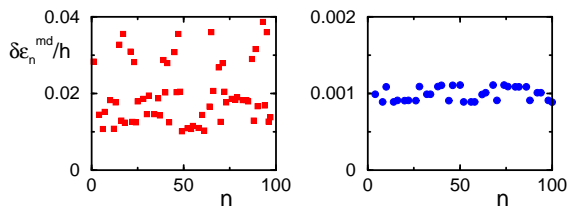}
\end{center}
\caption{ The low-energy parts of the two-particle energy differences
$\delta \varepsilon_n^{\rm md}$ (\protect\ref{band_modified}) for the
transitions with $\varkappa \leq 4$ in which one of the involved particles
is on the $n$th site ($n> 2$). The left and right panels refer to the
energy sequence (\ref{sequence}), (\ref{correction_6}) with $\alpha =
0.25, \alpha^{\prime}=0.22$ and $\alpha =0.1, \alpha^{\prime}
=1/15$, respectively. In agreement with the analytical estimates,
the zero-energy gap $\delta_0/h\sim \alpha^3$.}
\label{fig:zero_energy}
\end{figure}

With the appropriately chosen $\alpha^{\prime}$, the condition on
$\alpha$ imposed by Eq.~(\ref{gap_condition}) takes the form
\begin{equation}
\label{alpha_below}
\alpha^3 \gg 2(J/2h)^3\Delta.
\end{equation}
It can be seen that, when Eq.~(\ref{gap_condition}) holds, not only is
the many-particle renormalization of site energies small, but also the
matrix elements of $\varkappa\leq 4$ transitions turn out to be
smaller than $\delta_0$. For given $\alpha$ and $J/h$, the condition
(\ref{alpha_below}) sets the upper limit on the dimensionless
parameter of the particle-particle interaction $\Delta$.

We have checked numerically that the gap $\delta_0$ persists for a
much longer chain than shown in Fig.~\ref{fig:zero_energy}, with $n$
from 2 to 1000. This is sufficient to prove that the results apply to
an infinite chain, cf. Ref.~\cite{DISS04}. Indeed, in
$\varepsilon_n^{\rm md}$ the terms $\propto \alpha^{\prime}$ are
repeated periodically with period 6, whereas the terms
$\propto\alpha^q$ with different $q$ are repeated with period
$2(q+1)$. Therefore all combinations of terms $\alpha^0, \alpha^1,
\ldots, \alpha^q$ with $q\leq 6$ as well as the terms
$\alpha^{\prime}$ are repeated periodically with period equal to twice
the least common multiple of all $q+1\leq 7$, which is
$2\times(3\times 4\times 5\times 7)= 840$. For $\alpha = 0.25$ and
$\alpha=0.1$ we have $\alpha^6 \approx 2\times 10^{-4}$ and $\alpha^6=
10^{-6}$, respectively. The contribution to $\varepsilon_n^{\rm md}/h$
of all terms of higher degree in $\alpha$ is much smaller than the gap
$\delta_0$. This means that the results on the gap for an infinite
chain will coincide with the results for a chain of 840 sites.

\subsection{High-energy gap}

We now discuss the gap in the two-particle transition energies
$\delta\varepsilon_n^{\rm md}$ centered at $h/2$. We will be interested in the
gap for all $n$ and for all transitions with $\varkappa \leq 4$. The gap is
determined by the widths of the energy bands of $\delta\varepsilon^{\rm md}$
centered at $\delta\varepsilon_n^{\rm md}=0$ and $\delta\varepsilon_n^{\rm
md}=h$ for the energy sequence (\ref{sequence}), (\ref{correction_6}). Because
the sequence is regular, the single-energy distribution has no tails and the
bands have sharp boundaries.

It is seen from Eqs.~(\ref{sequence}), (\ref{correction_6}) that, in a
two-particle transition $(k_4,k_3)\leftrightarrow (k_1,k_2)$, the
terms in $\varepsilon_{k_1},\ldots,
\varepsilon_{k_4}$ that are linear in $\alpha$ 
may all add up or pairwise compensate each
other. The sum of all of them can be equal therefore to $\pm 2\alpha
h, \pm \alpha h$, or 0. Similarly, the term $\alpha^{\prime}h/2$ can
be added or subtracted. Note that, for $\varkappa \leq 4$, only one
term $\propto \alpha^{\prime}$ can contribute to
$\delta\varepsilon_n^{\rm md}$. As a result, the width of the band of
$\delta\varepsilon_n^{\rm md}$ at $\delta\varepsilon_n^{\rm md} =0$ is $
h(2\alpha +\alpha^{\prime}/2)$, whereas the width of the band at
$\delta\varepsilon_n^{\rm md}=h$ is $h(4\alpha +\alpha^{\prime})$.

The changes of the site energies in 3-particle transitions with
$\varkappa\leq 4$
\[|\varepsilon_{k_1}^{\rm md}+
\varepsilon_{k_2}^{\rm md}+
\varepsilon_{k_3}^{\rm md}-\varepsilon_{k_4}^{\rm md}-\varepsilon_{k_5}^{\rm md}-\varepsilon_{k_6}^{\rm md}|\]
also form bands centered at $0,\,h$ and $2h$. To leading order in
$\alpha, \alpha^{\prime}$, the bands at energies $0$ and $h$ are
bounded by $(2\alpha+\alpha^{\prime})h$ and $h(1 \pm 3\alpha \pm
\alpha^{\prime})$. Interestingly, four-particle transitions with
$\varkappa \leq 4$ do not change these bounds \cite{DS_to_be}.

>From this analysis, the gap in the changes of site energies of all
$\varkappa\leq 4$ transitions $(\delta\varepsilon)_{\varkappa\leq 4}$
is given by the condition
\begin{equation}
\label{gap_limits}
2\alpha + \alpha^{\prime} < h^{-1}(\delta\varepsilon)_{\varkappa\leq
4} < 1 - 3\alpha - \alpha^{\prime}.
\end{equation}
Higher-order terms in $\alpha$ can be taken into account by replacing
$\alpha \to \alpha/(1-\alpha)$ in Eq.~(\ref{gap_limits}). A direct
analysis of the terms $\sim \alpha^2$ shows that this gives a stronger
inequality than what actually follows from Eqs.~(\ref{sequence}),
(\ref{correction_6}).

The conditions for eliminating  resonant transitions with
$\varkappa\leq 4$ in an infinite chain are given by
Eq.~(\ref{gap_limits}) in which $h^{-1}(\delta\varepsilon)_{\varkappa
\leq 4}$ is replaced by $J\Delta$ and $2J\Delta$.  The distance from
$J\Delta, 2J\Delta$ to the band edges should significantly exceed the
transition matrix elements and the energy renormalization $J\Delta
(J/2h)^2$; these quantities are small compared to $\alpha h$.

\begin{figure}[h]
\begin{center}
\includegraphics[width=3.0in]{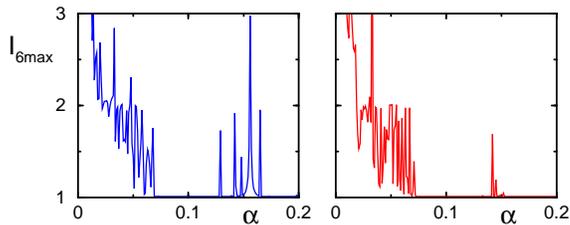}
\end{center}
\caption{The maximal inverse participation ratio $I_{6\,\max}$ for a system of
6 particles on a 12-site chain. The site energies $\varepsilon_n^{\rm md}$ are
given by Eqs.~(\ref{sequence}), (\ref{correction_6}) with
$\alpha^{\prime}=1/15$. The left and right panels corresponds to the sections
of the chain with $415\leq n \leq 426$ and $600\leq n\leq 611$, respectively.
The coupling parameter is $\Delta = 9$, and $h/J=30$. A broad region of
$\alpha$ centered at $\alpha=0.1$ where the IPR is extremely close to 1 is seen
in all sections of the chain that were tested \cite{DS_to_be}.}
\label{fig:IPR_strong}
\end{figure}

The inequalities (\ref{alpha_below}), (\ref{gap_limits}) show that the
range of the parameters $h/J,\Delta$, and $\alpha$ where two-particle
resonances with $\varkappa \leq 4$ are eliminated is fairly large. One
may expect that, in this range, stationary many-particle states will
be mostly confined to their sites, at least for a not too long section
of the chain. This can be probed by calculating the many-particle IPR
for a strongly coupled system. The results of these calculations for
modified energy sequence (\ref{correction_6}) are shown in
Fig.~\ref{fig:IPR_strong}. It is seen from this figure that, in a
broad range of $\alpha$ where the localization lifetime is expected
from Eqs.~((\ref{alpha_below}), (\ref{gap_limits} to be long, the IPR
is extremely close to one. This shows that all stationary 
many-particle states are strongly confined.

\section{Conclusions}

In this paper we studied strong single- and many-particle
confinement. We have shown that the previously proposed \cite{DISS04}
energy sequence (\ref{sequence}) leads to strong confinement of all
single-particle states. The decay of the wave functions is
quasi-exponential in both directions in the chain, with the same
bounds on the decay length.

We have also considered the problem of many-particle confinement. Here we
studied the localization lifetime $t_{\rm loc}$ during which all particles in
an infinite chain remain on their sites. Of central interest was an extension
of the results on long localization lifetime \cite{DISS04} to the case of
strong particle-particle coupling, $\Delta\gg 1$. We map the problem of
many-particle transitions onto the problem of resonant scattering. A natural
characteristic in the analysis of scattering is the difference between the
initial and final site energies $\delta\varepsilon$. Large $t_{\rm loc}$
compared to the reciprocal hopping integral $J^{-1}$ was obtained earlier
\cite{DISS04} when the zero-energy gap $\delta_0$ in $\delta\varepsilon$ for
all transitions with up to 4 single-particle moves exceeded twice the coupling
energy $2J\Delta$. For $\Delta \gg 1$ this would require a very large width of
the band of site energies.

We have shown that, for large $\Delta$, a better option may be to {\it
decrease} the bandwidths of the site energies of second neighbors in
the chain, which leads to separation and narrowing of bands of
$\delta\varepsilon$. Then, if $J\Delta$ and $2J\Delta$ are within the
interband gaps of $\delta\varepsilon$, resonant transitions are
eliminated.  As a result, the lifetime $t_{\rm loc}$ scales as a high
power of the reciprocal hopping integral, $t_{\rm loc}\propto
(J\Delta)^{-1}(2\alpha h/J)^6$ for transitions that involve one
particle-particle collision. The overall bandwidth of site energies
required for obtaining a long localization lifetime is parametrically
smaller than the one given by the condition $\delta_0 > 2J\Delta$. The
major limitation on the bandwidth is that it should substantially
exceed the coupling energy. The results bear on many-electron
conductivity in condensed-matter systems and on quantum computing with
perpetually coupled qubits. They demonstrate the possibility to avoid
delocalization of excitations in a QC without refocusing.

\ack
We acknowledge support by the Institute for Quantum Sciences at
Michigan State University and by the NSF through
grant No. ITR-0085922.

\end{document}